\documentclass[a4paper,12pt]{article}
\usepackage[margin=2.5cm]{geometry}
\usepackage{graphicx, caption, subcaption,lipsum}
\usepackage{amsmath}
\usepackage{newtxtext,newtxmath} 
\DeclareCaptionFont{myfontsize}{\fontsize{10}{11}\selectfont}
\title{Absence of skewness in the voltage fluctuations of a tunnel junction in the quantum regime}

\author{
    Clovis Farley$^1$ and Bertrand Reulet$^1$ \\
    \small $^1$Département de physique, Institut Quantique, Université de Sherbrooke, Sherbrooke QC, Canada \\
    \small \texttt{clovis.farley@usherbrooke.ca} \\
    \small \texttt{bertrand.reulet@usherbrooke.ca}
}

\date{}

\begin{document}

ICNF-2025 \phantom{xxxxxxxxxxxxxxxxxxxxxxxxxxxxxxxxxxxxxx} June 17-20, 2025 – Taormina, Italy

{\let\newpage\relax\maketitle}

\fontsize{11}{12}\selectfont

\textbf{Abstract:}
Current fluctuations in a tunnel junction have a remarkable property: On the one hand, their variance corresponds to vacuum fluctuations at low voltage bias $V$, when the electron energy $eV$ is smaller than the photon detection energy $hf$. On the other hand, their skewness, i.e. their third moment, is frequency independent, equal to $e^2I$ as if electron transport were simply Poissonian. We address the following question: Could it be that at low voltage, the vacuum fluctuations generated by the junction have a finite skewness, i.e. that the junction generates skewed vacuum ? To answer this question we calculate the effect of an arbitrary electromagnetic environment at zero temperature and show that the bispectrum of third moment of voltage fluctuations of any quantum conductor is always zero at frequencies larger than the voltage. We also show experimental data on tunnel junctions in the quantum regime that agree with our calculation.


\vspace{0.5 cm}
\fontsize{12}{14}\selectfont

\textbf{Introduction.}

The third moment of current/voltage fluctuations in an electrical device is known to be dependent on the electromagnetic environment of the sample, i.e. the circuit to which it is connected \cite{Beenakker2003,Reulet2003,Gershon2008}. This occurs as soon as the variance of current/voltage fluctuations depends on the bias voltage of the device, i.e. for all components but a linear resistor. This comes from two factors: i) external voltage fluctuations modulate the noise of the device, and ii) current fluctuations generated by the device generate voltage fluctuations on the device itself through the external impedance to which the device is connected. These contributions add to the intrinsic skewness of the component.

A general problem of quantum noise of an electronic conductor is to understand how electronic fluctuations imprint their statistical properties in the electromagnetic radiation they generate. In the case of the variance of fluctuations, the link between current fluctuations and electromagnetic power detected at frequency $f$ is well understood and depends on the detection scheme (photon detector, voltage detector, etc.)\cite{Intro_QNoise}. In the case of a classical detector, the averaged square voltage measured at frequency $f$ is proportional to $n(f)+\frac12$ with $n(f)$ the number of photons per second emitted in the detection bandwidth. The $1/2$ background corresponds to vacuum fluctuations. If electrons are given an energy $eV<hf$ they cannot emit photons so $n(f)=0$ and the detected noise is that of vacuum fluctuations. 

Here we want to probe what happens to the bispectrum $\langle v(f_1)v(f_2)v(f_3)\rangle$ in the limit where all frequencies $f_i$ are greater than $eV/h$, i.e. in the full quantum regime where the variance of the noise is that of vacuum at all frequencies $f_1$, $f_2$ and $f_3$. Could it be that the vacuum generated by a quantum conductor at low bias is skewed ? Or is it that the environmental contributions always conspire to make the skewness vanish in this regime ?

\vspace{0.3cm}

\textbf{Theory.} We consider the general case depicted in Fig. 1: a quantum conductor of resistance $R$ is connected to a perfect amplifier via an arbitrary scattering matrix $S$. The perfect amplifier is made of a circulator with a load $Z_0$ at zero temperature, followed by a real amplifier. With such a scheme, the fluctuations generated by both the conductor and the vacuum fluctuations generated by the circulator load are partially reflected by the environment and partially transmitted to the amplifier. All parasitic capacitances, inductances, delay lines, etc. are incorporated in the $S$-matrix of the environment. The $S$-matrix is chosen with port 1 (the sample) having an impedance $R$ and port 2 (the amplifier) an impedance $Z_0$. $S_{ij}$ represents the scattering amplitude from port $j$ to port $i$.

\begin{figure}
    \centering
    \includegraphics[width=0.5\linewidth]{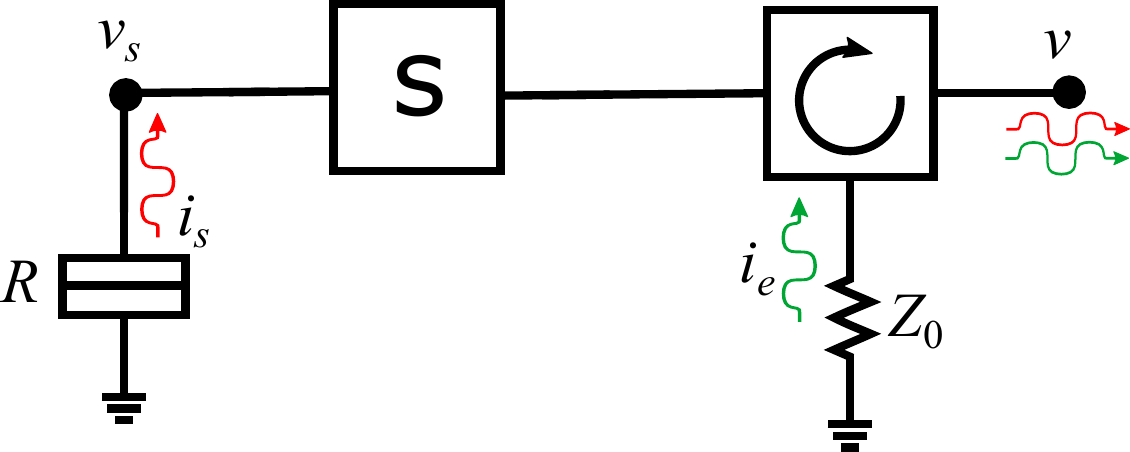}
    \caption{Detection scheme. Green arrows represent the vacuum fluctuations generated by the load $Z_0$. Red arrows represent the noise generated by the sample at frequencies $f_1$ and $f_2$.}
    \label{fig:scheme}
\end{figure}

According to the circuit of Fig. \ref{fig:scheme}, the voltage detected by the amplifier is given by:
\begin{equation}
    v(f)=\frac12S_{22}(f)Z_0i_e(f)+\frac12S_{21}(f)Ri_s(f)
\end{equation}
Here $i_e$ and $i_s$ represent the current noise generated by the load and the sample respectively. 
The voltage fluctuations across the sample are given by:
\begin{equation}
    v_s(f)=\frac12S_{12}(f) Z_0i_e(f)+\frac12\big( 1+S_{11}(f) \big) Ri_s(f)
    \label{eq:v_s}
\end{equation}
Note that the reflection coefficient $\Gamma=(R-Z_0)/(R+Z_0)$ on the sample from an impedance $Z_0$ does not appear thanks to the choice of the port 1 of the matrix $S$ being $R$, not $Z_0$. 
We start by calculating the power per unit bandwidth detected at frequency $f$:
\begin{equation}
    P_0=\frac1{Z_0}\langle |v(f)|^2\rangle=\frac14|S_{22}(f)|^2Z_0N_e(f)+\frac14|S_{21}(f)|^2\frac {R^2}{Z_0}N_s(f)
\end{equation}
Here $N_e(f)=\langle i_e(f)i_e(-f)\rangle$ and $N_s(f)=\langle i_s(f)i_s(-f)\rangle$ are respectively the spectral densities of the current noise generated by the load and the sample. At $T=0$, the load generates vacuum fluctuations, $N_e(f)=h|f|/Z_0$. For the sample, we first focus on a tunnel junction, which when biased at voltage $V$ generates current fluctuations of spectral density:
\begin{equation}
    N_s(f)=\frac{|eV-hf|+|eV+hf|}{2R}
\end{equation}
For $e|V|<h|f|$, $N_s(f)=h|f|/R$ (this is valid for any sample, not just the tunnel junction). In this limit, one has:
\begin{equation}
    P_0=\frac{hf}4\Big(|S_{22}(f)|^2+|S_{21}(f)|^2\frac {R}{Z_0}\Big)
\end{equation}
Thanks to the unitarity of the $S$ matrix, the quantity within parenthesis is 1, so $P_0$ corresponds to vacuum fluctuations, part of which coming from the sample, the rest from the load. The factor $R/Z_0$ comes from the choice of the impedance of the ports of the scattering matrix $S$ \cite{Pozar}.

We now consider the skewness $\langle v(f_1)v(f_2)v(f_3)\rangle$. Since the frequencies must obey $f_1+f_2+f_3=0$ in the absence of ac excitation, we take $0<f_1\leq f_2 < -f_3=f_1+f_2$. The current fluctuations of the sample and the load are uncorrelated, however the noise of the sample can be correlated to the voltage across itself, so correlators involved in the skewness are:
\begin{equation}
    \langle v(f_1)v(f_2)v(f_3)\rangle=   \langle j_{s,1}j_{s,2}j_{s,3}\rangle
    +\langle j_{s,1}j_{s,2}j_{e,3}\rangle
    +\langle j_{s,1}j_{e,2}j_{s,3}\rangle
    +\langle j_{e,1}j_{s,2}j_{s,3}\rangle
    \label{eq:fit1}
\end{equation}
with $j_{s,k}=(R/2)S_{21}(f_k)i_s(f_k)$ and $j_{e,k}=(Z_0/2)S_{22}(f_k)i_e(f_k)$.
Terms of the form $\langle j_s^2j_e\rangle$ correspond to the modulation of the sample noise by the external vacuum fluctuations and are calculated using:
\begin{equation}
    \langle i_s(f_i)i_s(f_j)i_e(f_k)\rangle=\frac{d\langle i_s(f_i)i_s(f_j)\rangle}{dv_s(-f_k)} \langle i_e(f_k)v_s(-f_k)\rangle
    \label{eq:fit2}
\end{equation}
with $\langle i_e(f)v_s(-f)\rangle=\frac12 S_{12}(-f)Z_0N_e(f)$. 
The quantity:
\begin{equation}
    \chi_{ijk}=\frac{d\langle i_s(f_i)i_s(f_j)\rangle}{dv_s(-f_k)}
\end{equation}
is the noise susceptibility, which measures the linear response of the noise to a voltage excitation across the sample at frequency $-f_k$. It has been studied theoretically and experimentally \cite{Gabelli2008, Gabelli_SPIE,Qtape,Gabelli2009,Farley2023}. At zero temperatures, it reduces, for a tunnel junction, to:
\begin{equation}
    \chi_{ijk}=\frac{e}{2h|f_k|R}[|hf_i+eV|-|hf_i-eV|+|hf_j+eV|-|hf_j-eV|]
\end{equation}
In the low voltage limit $e|V|<hf_1$,
\begin{equation}
    \chi_{231}=\chi_{312}=0
\end{equation}
i.e., excitations at frequencies $f_1$ or $f_2$ do not modulate the noise at frequency $f_3$.
\begin{equation}
    \chi_{123}=\frac{2e^2I}{h|f_3|}
\end{equation}
Thus the contribution of the environmental noise on the skewness bispectrum $\langle v(f_1)v(f_2)v(f_3)\rangle$ at low voltage bias is:
\begin{equation}
    \frac{R^2Z_0}{8}S_{21}(f_1)S_{21}(f_2)S_{22}(f_3)S_{12}(-f_3)e^2I
    \label{eq:contribution_env}
\end{equation}
This contribution stems from voltage fluctuations across the sample due to external noise. The noise of the sample itself contributes also to these fluctuations, leading to the feedback term, which is similar with $\langle i_s(f)v_s(-f)\rangle=\frac12 \big( 1+S_{11}(-f) \big) R N_s(f)$,   see Eq.(\ref{eq:v_s}). These terms contribute to the skewness as:
\begin{equation}
    \frac{R^3}{8}S_{21}(f_1)S_{21}(f_2)S_{21}(f_3)\big( 1+S_{11}(-f_3) \big)e^2I
    \label{eq:contribution_feedback}
\end{equation}
Finally the intrinsic electronic skewness contributes to the same quantity by:
\begin{equation}
    -\frac{R^3}{8}S_{21}(f_1)S_{21}(f_2)S_{21}(f_3)e^2I
    \label{eq:contribution_intrinsic}
\end{equation}
Adding the three contributions leads to the prefactor:
\begin{equation}
    S_{21}(f_3)S_{11}(-f_3)+S_{22}(f_3)S_{12}(-f_3)\frac{Z_0}{R}=0
\end{equation}
The unitarity of the $S$ matrix ensures that this vanishes, taking into account the impedances of the port chosen for the scattering matrix \cite{Pozar}. Thus $\langle v(f_1)v(f_2)v(f_3)\rangle=0$ at low voltage: a tunnel junction does not generate skewed vacuum. The case of an arbitrary quantum conductor is more subtle: its noise at low voltage still reduces to vacuum fluctuations but its noise susceptibility is multiplied by a Fano factor $F_2$ which is frequency independent. In contrast, its intrinsic skewness is affected by another factor $F_3(f_1,f_2)$ which depends on voltage and frequency \cite{Galaktionov2003,Salo2006,Qtape}. In such a case, Eqs.(\ref{eq:contribution_env}) and (\ref{eq:contribution_feedback}) should be multiplied by $F_2$ and Eq. (\ref{eq:contribution_intrinsic}) by $F_3$. We find:
\begin{equation}
    \langle (v(f_1)v(f_2)v(f_3)\rangle=\frac{R^3}{8} S_{21}(f_1)S_{21}(f_2)S_{21}(f_3)(F_2-F_3)e^2I
\end{equation}
According to \cite{Qtape}, $F_3(f_1,f_2)=F_2$ for $eV<hf_{1,2}$ so we find again that $\langle (v(f_1)v(f_2)v(f_3)\rangle=0$ at low voltage: the nonzero skewness of current fluctuations in a coherent conductor never translates into skewed vacuum. It is noteworthy that the equality $F_2=F_3$ comes from a calculation involving electrons only while it acquires a clear physical meaning when one considers the electromagnetic field radiated by the sample being vacuum at low voltage.

\vspace{0.3cm}

\begin{figure}
    \centering
    \includegraphics[width=0.5\linewidth]{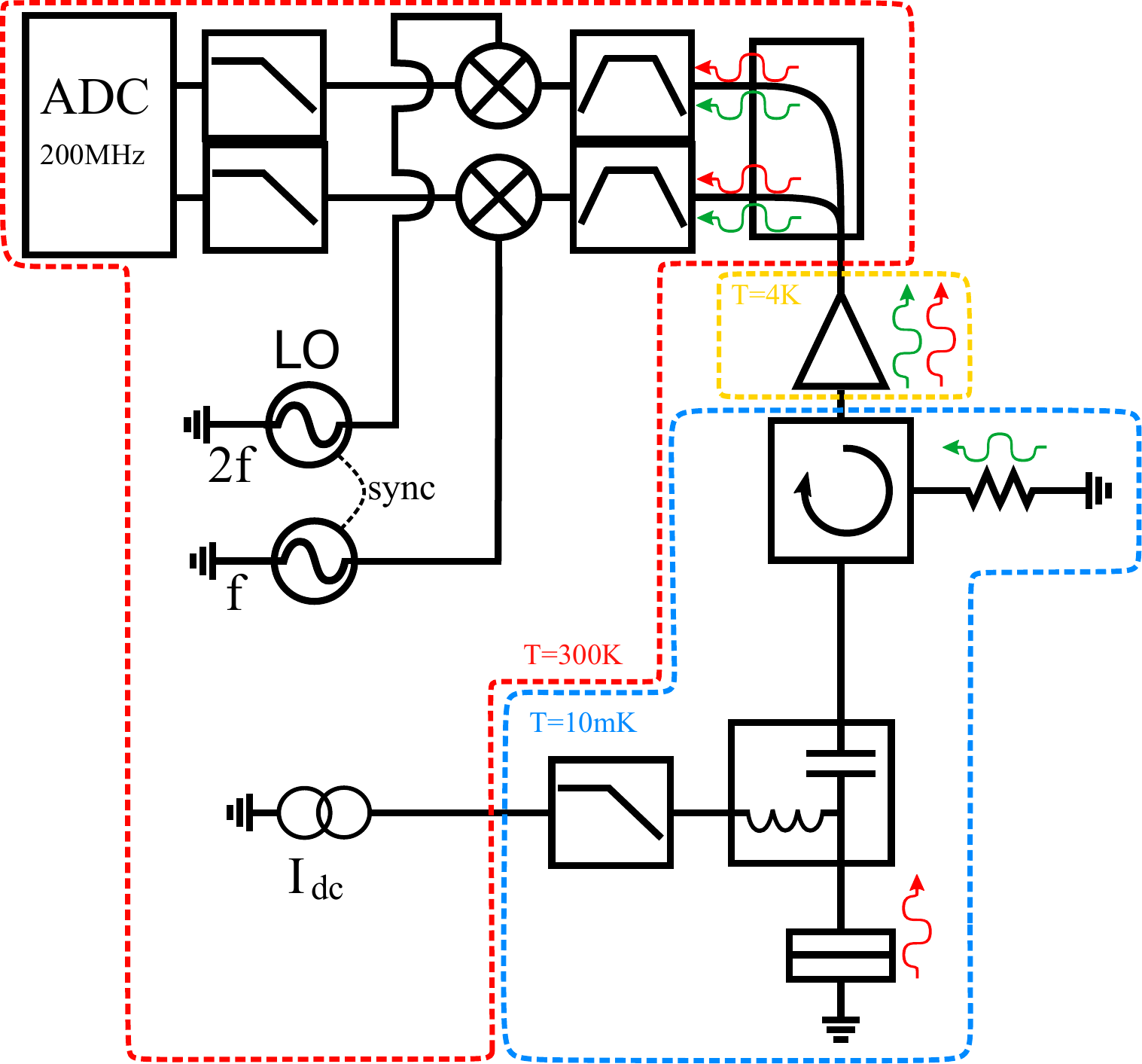}
    \caption{Experimental setup.}
    \label{fig:setup}
\end{figure}

\textbf{Experiment.} We have performed measurements of the skewness of voltage fluctuations generated by a tunnel junction measured at frequencies $f_1=f_2=f$, $f_3=-2f$ with $f=5.05$ GHz. We have measured a tunnel junction of resistance $R=93$ $\Omega$, i.e. relatively matched to the detection setup ($Z_0=50$ $\Omega$) as well as a much more resistive sample of resistance $R=505$ $\Omega$. For the latter, the detection is a good ammeter ($Z_0\ll R$) for which environmental corrections are small. In this limit, the intrinsic contribution $e^2I$ is naively expected to dominate, possibly leading to nonzero third moment at low voltage, even though our calculation predicts the contrary. 

The experimental setup is shown in Fig. 2. The detection is similar to that of \cite{Farley2025}: the sample, a tunnel junction fabricated using standard techniques of photo-lithography, is DC biased via a bias tee. The fluctuations it generates are amplified and separated in two branches at $f$ and $2f$, which are separately downconverted, by mixing at frequencies $f$ and $2f$, into slow-varying signals $X_1$ and $X_2$ respectively (of bandwidth 120 MHz). We record these signals and compute the time-averaged correlator $\langle X_1^2X_2\rangle$, which is proportional to the real part of $\langle (v^2(f)v(-2f)\rangle$, noted $\langle v^3 \rangle$ for brevity. Rotating the phase of the local oscillators at $2f$ allows to measure the other quadrature, $Y_2$, from which we compute the correlator $\langle X_1^2Y_2\rangle$, proportional to the imaginary part of the skewness. Circulators between the cryogenic amplifier and the sample ensure that the effective load seen by the sample is at an effective temperature $T \ll hf/k_B$. In our setup, the noise generated by the load that is detected, which is measured by $S_{22}$, comes from its reflection on the sample, i.e $S_{22}=\Gamma$. Similarly, $S_{11}$ comes from the mismatch between the sample and the 50 $\Omega$ coax cable connected to it, i.e. $S_{11}=\Gamma$. The electron temperature is determined by fitting the voltage dependence of the noise $N_e$ generated by the sample at $f$ and $2f$. We find $T\simeq$20 mK , i.e. $k_BT/(hf)\simeq0.08$.

\begin{figure}
    \centering
    \includegraphics[width=0.9\linewidth]{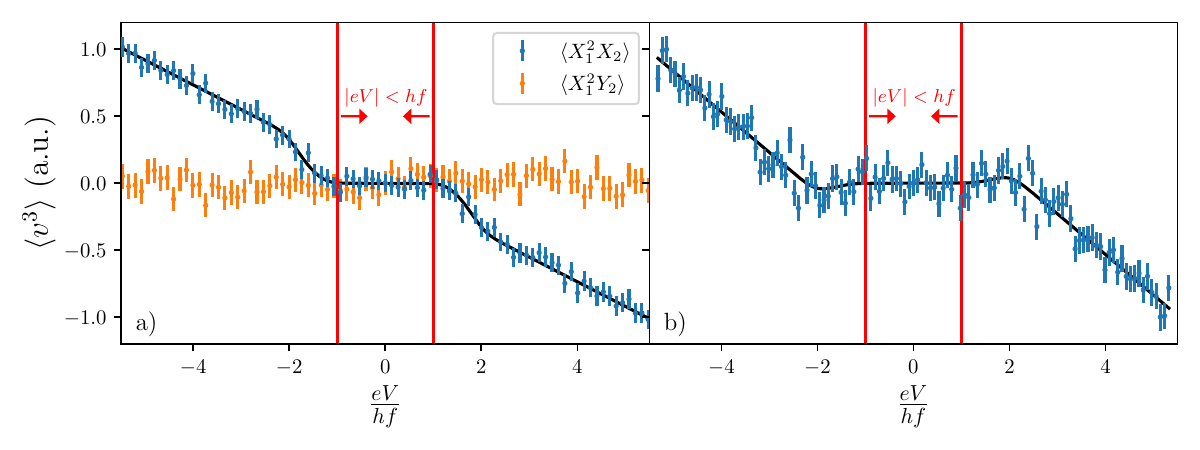}
    \caption{Skewness of the voltage fluctuations for a tunnel junction of resistance 93 $\Omega$ (a) and 505 $\Omega$ (b). Symbols are experimental data, solid lines theoretical predictions.}
    \label{fig:manip}
\end{figure}

Losses and amplification between the sample and the digitizer affect $S_{21}$. This results in an overall unknown gain, which is the only fitting parameter. Losses between the load of the last circulator and the sample are accompanied with re-emission of noise at ultra-low temperature, so that the sample sees a source of zero-temperature noise anyway.

We show in Fig. 3(a) the result of the measurement for the sample of resistance $93$ $\Omega$ together with a theoretical fit, which corresponds to $\Gamma=0.30$. 
We show in Fig. 3(b) the result of the measurement for the sample of resistance $505$ $\Omega$ together with a theoretical fit, which corresponds to $\Gamma=0.82$. The fits are performed using Eqs. (\ref{eq:fit1}) and (\ref{eq:fit2}) with the expressions of the noise susceptibility given in \cite{Gabelli_SPIE,Qtape,Farley2023}, with one free parameter, the overall gain of the setup. In both figures theory and experiment agree very well. We always observe that the skewness vanishes at voltage smaller than $hf/e$, in agreement with our calculations. In a previous high frequency experiment, the correlator $\langle v(\epsilon)v(f-\epsilon)v(-f)\rangle$ has been measured, with $\epsilon<400$ MHz and $f=6$ GHz. \cite{Gabelli2013}. Despite the fact that this experiment was not fully in the quantum regime, since $h\epsilon>k_BT$, it was observed that the skewness of the detected voltage vanished at low dc bias $V<hf/e$. This was attributed to a good impedance match between the sample and the setup (indeed our result that the low bias skewness vanishes irrespective of the impedance of the sample does not apply to this experiment).

\textbf{Perspectives.} We have shown theoretically that the skewness of the electromagnetic field radiated by a quantum conductor at low voltage always vanishes. We have confirmed this result experimentally in the case of a tunnel junction. While this makes sense physically, it demands for a physical interpretation in terms of detection and raises the question of what would happen for another type of detection, that e.g. involves correlation between a voltage amplifier and a photon counter, or a system with discrete levels \cite{Brosco2006,Heikkila2007,Gabelli2009}. 

\textbf{Acknowledgments.} We thank Julien Gabelli and Jérôme Estève for fruitful discussions. We are very grateful to Edouard Pinsolle for providing us with the tunnel junctions and to Gabriel Laliberté and Christian Lupien for their technical help. This work was supported by the Canada Research Chair program, the NSERC, the Canada First Research Excellence Fund, the FRQNT, and the Canada Foundation for Innovation.

\bibliography{biblio}
\bibliographystyle{unsrt}

\end{document}